\documentclass[aps,prl,preprint,groupedaddress]{revtex4-2}

\usepackage{graphicx}
\usepackage{float}
\usepackage{xcolor}

\begin{document}


\title{Layer number and stacking order-dependent thermal transport in molybdenum disulfide with sulfur vacancies}


\author{Ranjuna M K}
\email[]{ranjuna.mk@gmail.com}
\author{Jayakumar Balakrishnan}
\email[]{jayakumar@iitpkd.ac.in}
\affiliation{Department of Physics, Indian Institute of Technology Palakkad, Palakkad-678623, Kerala, India}


\date{\today}
\begin{abstract}
Recent theoretical works on two-dimensional molybdenum disulfide, MoS$_2$, with sulfur vacancies predict that the suppression of thermal transport in MoS$_2$ by point defects is more prominent in monolayers and becomes negligible as layer number increases. Here, we investigate experimentally the thermal transport properties of two-dimensional molybdenum disulfide crystals with inherent sulfur vacancies. We study the first-order temperature coefficients of interlayer and intralayer Raman modes of MoS$_2$ crystals with different layer numbers and stacking orders. The in-plane thermal conductivity ($\kappa$) and total interface conductance per unit area ($ g $) across the 2D material-substrate interface of mono-, bi- and tri-layer MoS$_2$ samples are measured using the micro-Raman thermometry. Our results clearly demonstrate that the thermal conductivity  is significantly suppressed by sulfur vacancies in monolayer MoS$_2$. However, this reduction in $\kappa$ becomes less evident as the layer number increases, confirming the theoretical predictions. No significant variation is observed in the $\kappa$ and $ g $ values of 2H and 3R stacked bilayer MoS$_2$ samples. 

\end{abstract}

\maketitle

\section{Introduction}

Two-dimensional transition metal dichalcogenides (TMDs), an advancing class of layered materials, have recently become a fertile ground for investigating fundamental properties and emergent device applications. Among these, 2D crystals of molybdenum disulfide (MoS$_2$) have gained significant importance due to their unique properties, such as direct bandgap in monolayer\cite{mak2010atomically}, high carrier mobility, significant light-matter interactions\cite{liu2015strong}, reasonable spin-orbit interactions\cite{xiao2012coupled}, good valley selectivity\cite{mak2012control,zeng2012valley,chakrabarti2022enhancement} and high Seebeck coefficient\cite{buscema2013large,wu2014large,huang2016metallic}. This makes them potential materials for applications in electronics\cite{radisavljevic2011single,chen2018tuning,cheng2014few}, optoelectronics\cite{fontana2013electron,pham2019mos2,bernardi2013extraordinary}, Valleytronics\cite{zeng2012valley,mak2012control,sun2019separation} and energy harvesting.\cite{huang2016metallic,xu2022frenkel} However, the number and stacking order of layers and the defects and impurities in MoS$_2$ can significantly impact its properties and the device's performance.\cite{qiu2013hopping} 

Monolayer MoS$_2$ comprises the hexagonal arrangement of molybdenum (Mo) and sulfur (S) atoms sandwiched to form an S-Mo-S structure. These monolayers can be stacked in different configurations to form different polytypes of multilayer MoS$_2$. Previous studies on the most common configurations, the 2H (hexagonal) and 3R (rhombohedral) phases, reveal that stacking order is vital in various applications like piezoelectricity, nonlinear optics, and catalytic activity.\cite{toh20173r,shi20173r,hallil2022strong} Similarly, 
structural defects in MoS$_2$, such as vacancies, substitutions, dislocations, grain boundaries, and edges, can significantly influence its properties due to their impact on the atomic structure and electronic behavior.\cite{qiu2013hopping,chen2018tuning,zhou2013intrinsic,tongay2013defects} However, depending on the targeted application, their impact can be either detrimental or beneficial. For example, the mobility of carriers in electronic devices can get reduced by the collision with defects. In contrast, the presence of defects can enhance the performance of MoS$_2$ in various fields, including catalysis, energy storage, and sensing.\cite{xu2022frenkel}

Understanding the effect of stacking order and structural defects on thermal transport in MoS$_2$ is crucial for optimizing its thermal properties and thermoelectric applications and designing efficient thermal management strategies in various applications. Thermal conductivity of monolayer and multilayer MoS$_2$ in its suspended and substrate-bound form has been reported.\cite{yan2014thermal, sahoo2013temperature, zhang2015measurement,yuan2018nonmonotonic} However, most of these works does not account for the possibility of structural defects or difference in the stacking order of MoS$_2$ crystals. Recently there have been theoretical studies on the thermal transport in MoS$_2$ with various defects like grain boundaries, vacancies, and substitutions.\cite{ding2015manipulating,peng2016beyond, saha2016theoretical,wang2016remarkable,lin2019phonon, polanco2020defect,xu2022grain,gabourie2020reduced} A recent experimental work explored the phonon thermal transport in few-layer MoS$_2$ flakes with various point defect concentrations enabled by helium ion (He$^+$) irradiation\cite{zhao2021modification}. They observed that Mo vacancies significantly reduce the thermal conductivity of few-layer MoS$_2$ compared to S vacancies. Similarly, recent work on thermal transport in CVD-grown monolayer MoS$_2$ with grain boundaries also showed a lower thermal conductivity of the sample.\cite{yarali2017effects}  

 \begin{figure*}
 \includegraphics[width=0.98\textwidth]{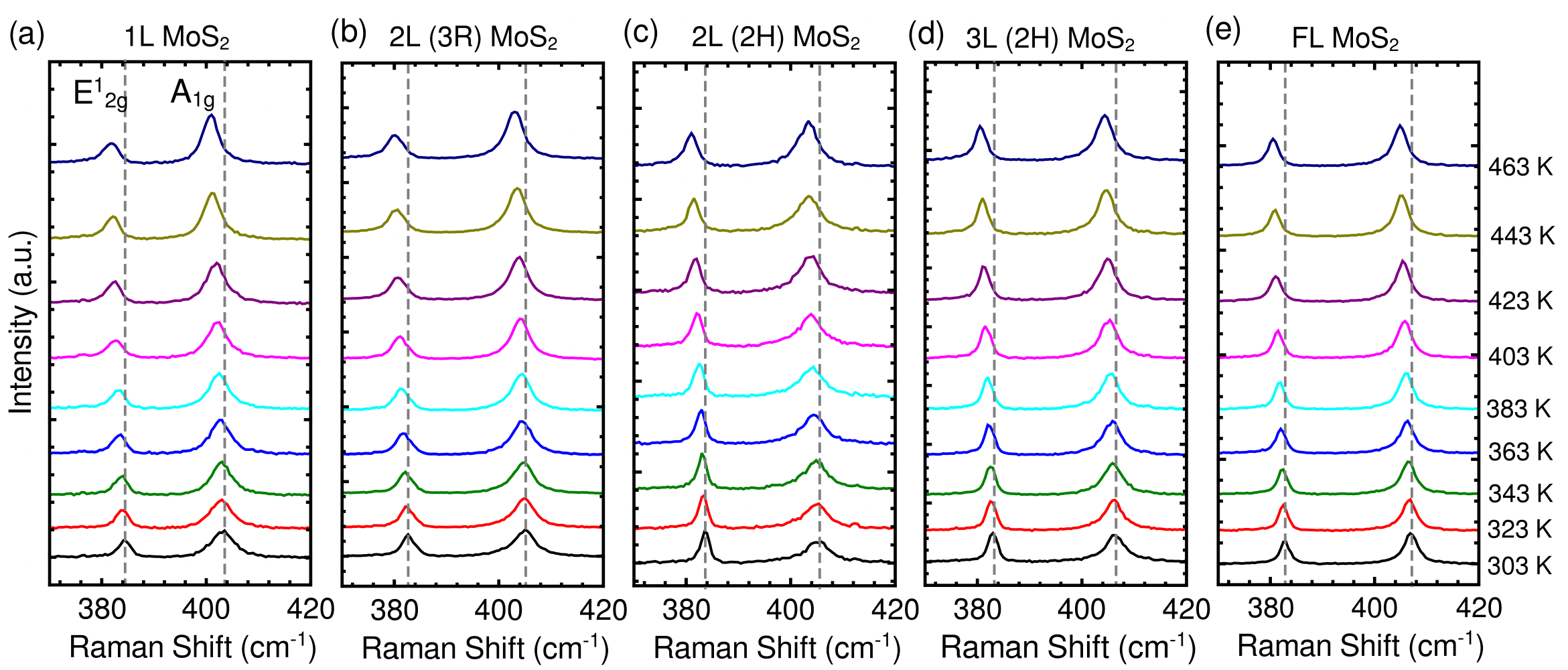}%
 \caption{\label{Figure1} (color online). The high frequency range of the Raman spectrum acquired at different temperatures using 532 nm excitation wavelength for (a) mono-, (b-c) bi-, (d) tri- and (e) few-layer MoS$_2$ crystals exfoliated on Si/SiO$_{2}$ substrate. The dashed (-) vertical lines indicates the position of E$^{1}_{2g}$ and A$_{1g}$ modes at 303 K. (f) The peak position of the  A$_{1g}$ and  E$^{1}_{2g}$ phonon modes at different temperatures for mono-, bi-, tri-, and few-layer MoS$_{2}$ crystals. The dashed (-) lines represent linear fit to data points.}
 \end{figure*}

In this article, we report a systematic study of thickness-dependent thermal transport in supported MoS$_2$ crystals with  sulfur vacancies, including the 2H and 3R stacked bilayers. The in-plane thermal conductivity ($\kappa$) and total interface conductance per unit area ($g$) across the 2D material-substrate interface are obtained using Raman thermometry. For this, in the first part of this work, we provide a detailed analysis of the first-order temperature coefficients for both intra- and inter-layer Raman modes. As per our understanding, this is the first report on the evolution of first-order temperature coefficients of low-frequency interlayer Raman modes of MoS$_2$ with layer number. In the second part, we utilize the first-order temperature coefficient of the Raman mode to calibrate sample temperature during laser heating and, consequently, determine the thermal transport parameters of the sample. Our experimental results show that the $\kappa$ decreases as the layer number increases, whereas the $g$ increases. Our results also demonstrate that the thermal conductivity is significantly suppressed by sulfur vacancies in monolayer MoS$_2$. However, this reduction in $\kappa$ becomes less evident as the layer number increases. Hence our results confirm the recent theoretical predictions that the suppression of thermal transport in MoS$_2$ by point defects is more prominent in monolayers and becomes negligible as layer number increases due to smaller density of sulfur vacancies. No significant variation is observed in the $\kappa$ and $ g $ values of 2H and 3R stacked bilayer MoS$_2$ samples.

\section{RESULTS AND DISCUSSIONS}

The 2D crystals of MoS$_2$ samples are prepared from bulk crystal by mechanical exfoliation and transferred on silicon wafers covered by 285 nm thick SiO$_2$. The numbers of MoS$_2$ layers are first identified by optical contrast and then confirmed by Raman spectroscopy. The stacking order of bi- and trilayer samples is determined using interlayer phonon modes in their Raman spectra. Previous works identified sulphur vacancy as the dominant category of defects in mechanically exfoliated MoS$_2$ samples.\cite{hong2015exploring} In our work, the exfoliated MoS$_2$ samples are annealed in an H$_2$-Ar environment for 3 hours at 350$ ^{\circ}$C. It has been shown that the annealing of MoS$_2$ crystals in H$_2$-Ar environment above 300$ ^{\circ}$C resulting in the creation of additional sulfur vacancies.\cite{zhu2023room} The stoichiometry of the molybdenum disulfide (Mo:S ratio) is obtained in the range of 1:1.70 to 1:1.93 from the Scanning Electron Microscopy Energy Dispersive Spectroscopy (SEM-EDS) and in the range of 1:1.71 to 1:1.76 from the X-ray Photoelectron Spectroscopy (XPS) measurement which confirms the presence of sulfur vacancies in the sample.\cite{ranjuna2023high} The 15\% sulfur vacancies corresponding to stoichiometric ratio of 1:1.7 is considered as the upper limit since the exposed area covers sample edges. To confirm further, Transmission Electron Microscopy - Energy Dispersive Spectroscopy (TEM-EDS) analysis is performed, and an Mo:S ratio of 1:1.80 is obtained. Additional details on sample preparation and characterization can be found in Ref. \cite{ranjuna2023high} and the supporting information file.\cite{SuppMat} 

The thermal transport properties of 2D crystals of MoS$_2$ are measured using Raman thermometry.\cite{cai2010thermal} This method benefits from its non-contact and non-destructive nature and relatively simple implementation. The experiment involves the application of a localized heat source to the material and measuring the resulting temperature changes. In the context of opto-thermal Raman techniques, the first-order temperature coefficient can serve as a valuable tool for calibrating the average sample temperature when exposed to laser heating. Later a steady-state heat conduction model is used to extract the $\kappa$ and $g$ values of the 2D materials bound to a substrate. 

\begin{figure}[ht]
  \includegraphics[width=0.35\textwidth]{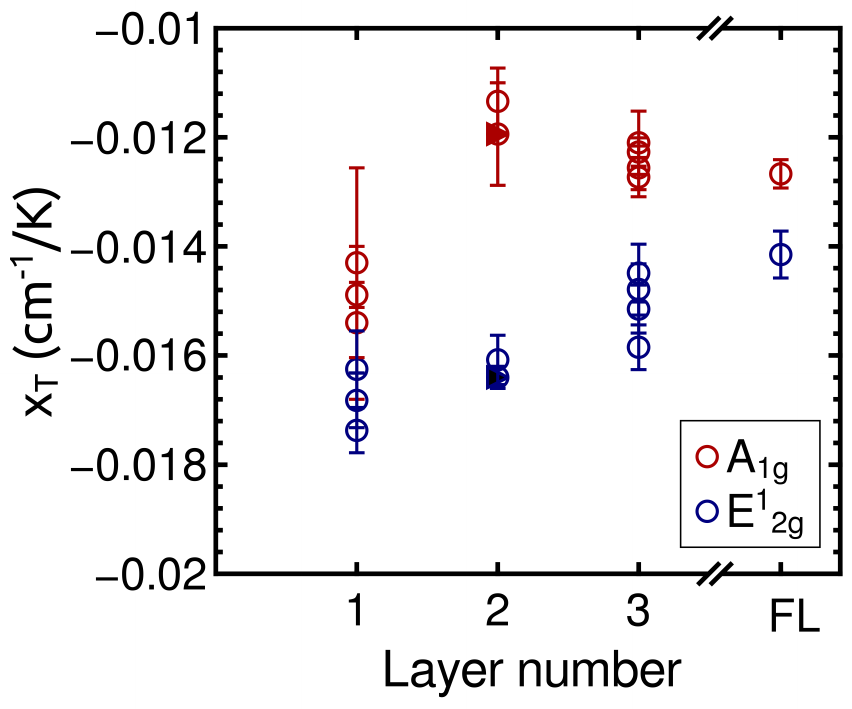}
  \caption{(color online). The value of first-order temperature coefficient of E$_{2g}^1$ and A$_{1g}$ modes for mono-, bi- tri- and few-layer MoS$_2$ crystals on Si/SiO$_2$ substrate. Different data points indicate different samples and the triangles indicate 3R stacked bilayer.} \label{Figure2}
\end{figure}

Initially, a low-power ($\sim 200 \mu W$) 532 nm laser beam is used to acquire the Raman spectrum of uniformly heated MoS$_2$ samples at different temperatures. The Figure \ref{Figure1} (a)-(e) represents the Raman spectra of MoS$_2$ samples recorded at different temperatures. The peaks around 384 cm$^{-1}$ and 406 cm$^{-1}$ correspond to the intralayer E$^1_{2g}$ and A$_{1g}$ modes, respectively. The E$_{2g}^1$ mode originates from the in-plane vibrations of the molybdenum (Mo) and sulfur (S) atoms, and the A$_{1g}$ mode originates from the out-of-plane vibrations of S atoms.\cite{li2012bulk} In all samples, the Raman mode frequency redshifts with increased temperature; as shown in Figure \ref{Figure1} (a)-(e). This observed redshift in phonon frequency is mainly due to the contribution from thermal expansion and anharmonic coupling of phonon modes at higher temperatures.\cite{sahoo2013temperature} The Raman mode frequencies corresponding to increase in temperature are fitted with a linear function ($ \omega (T) = \omega_0 + \chi_T \times T$) to determine the first-order temperature coefficient ($ \chi_{T}$) of each phonon mode. The comparison of first-order temperature coefficient of E$^1_{2g}$ and A$_{1g}$ modes of MoS$_2$ flakes with different layer number and stacking order is provided in the Figure \ref{Figure2}. For all layer numbers and stacking orders under study, the magnitude of $ \chi_{T}$ value of E$^1_{2g}$ mode is higher than that of A$_{1g}$ mode which agrees with previous reports.\cite{sahoo2013temperature, kim2021temperature} This difference in temperature coefficients between the E$^1_{2g}$ and A$_{1g}$ Raman modes in MoS$_2$ can be attributed to their different symmetries and vibrational characteristics. Since the E$^1_{2g}$ mode corresponds to the in-plane vibration of S and Mo atoms, it is particularly sensitive to the thermal expansion and anharmonic effects in the in-plane direction, contributing to its higher temperature coefficient. We observed that the $\chi_{T}$ values of the E$^1_{2g}$ mode exhibited a monotonous decrease in its magnitude as the layer number increased, whereas that of the A$_{1g}$ mode exhibited a sharp reduction from monolayer to bilayer and then increased with layer number.

  \begin{figure}
  \includegraphics[width=0.49\textwidth]{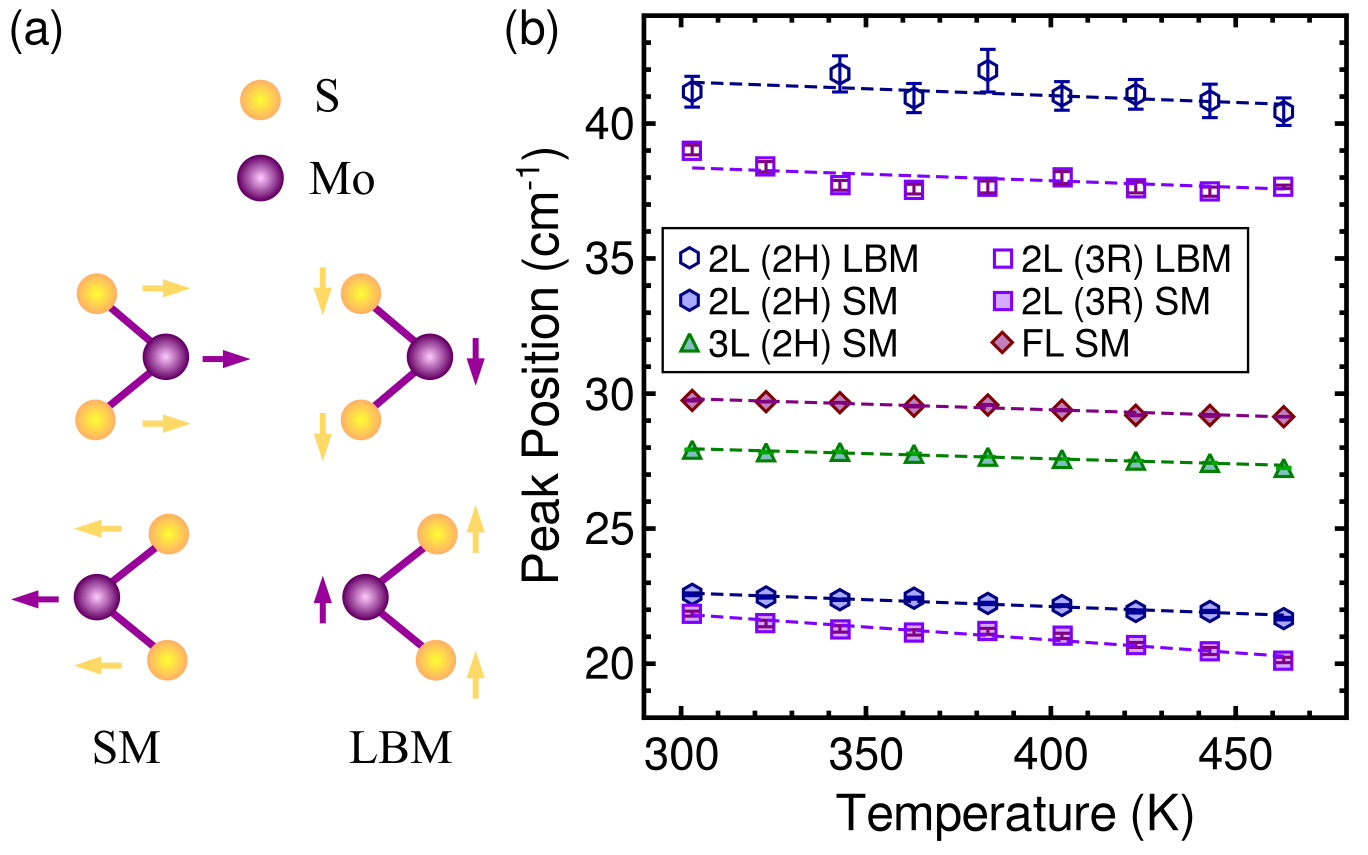}
 \caption{\label{Figure3} (color online). (a) The peak position of the layer breathing mode (LBM) and shear mode (SM) at different temperatures for bi-, tri-, and few-layer MoS$_{2}$ crystals exfoliated on Si/SiO$_{2}$ substrate. (X and Y axis in linear scale). The dashed (-) lines represent linear fit to data points. (b) Schematic of the interlayer vibrational modes of MoS$_2$}
 \end{figure}

Other than E$^1_{2g}$ and A$_{1g}$ modes, the Raman spectrum of MoS$_2$ samples contains interlayer phonon modes that appear in the low-frequency range (below 50 cm$^{-1}$). Shear mode (SM) and layer breathing mode (LBM) originate from in-plane and out-of-plane vibrations of both S and Mo atoms, respectively, as shown in Figure \ref{Figure3}(a). These modes are absent in the monolayer. The temperature coefficients of these interlayer modes in MoS$_2$ are not extensively studied as other Raman modes, primarily due to their relatively weaker intensity and the requisite of a more complex experimental setup with access to ultra-low frequencies and better spectral resolution.\cite{maity2018temperature,kim2021temperature} Here, we analyze the temperature dependence of low-frequency Raman modes of bi-, tri- and few-layer MoS$_2$ flakes. Figure \ref{Figure3}(b) displays the variation in the peak position of low-frequency modes as a function of temperature and the values of the first-order temperature coefficients are given in Table SI.\cite{SuppMat} Both shear and layer breathing modes show redshift due to anharmonicity and thermal expansion at higher temperatures.\cite{maity2018temperature} However, the $ \chi_T$ values of interlayer modes are much smaller than that of the high-frequency intralayer modes since these layers are coupled via weak Van der Waals force. The thermal expansion along the out-of-plane direction is negligible because the effect of temperature on the interlayer distance is considerably weak. Therefore, the contribution of crystal thermal expansion to the temperature coefficient can be ignored for the layer breathing mode.\cite{kim2021temperature} This justifies the small temperature coefficient value of layer breathing mode. In contrast, the shear mode is affected by thermal expansion, giving rise to a relatively larger temperature coefficient than the breathing mode. Additionally, the temperature coefficients of low-frequency modes can also depend on the interlayer interaction strength. The difference in the first-order temperature coefficient of the shear mode for 2H and 3R polytypes could be a manifestation of the difference in their interlayer coupling strength. The $ \chi _T$ value of the shear mode reduces their magnitude as the layer number increases. 

\begin{figure*}
 \includegraphics[width=\textwidth]{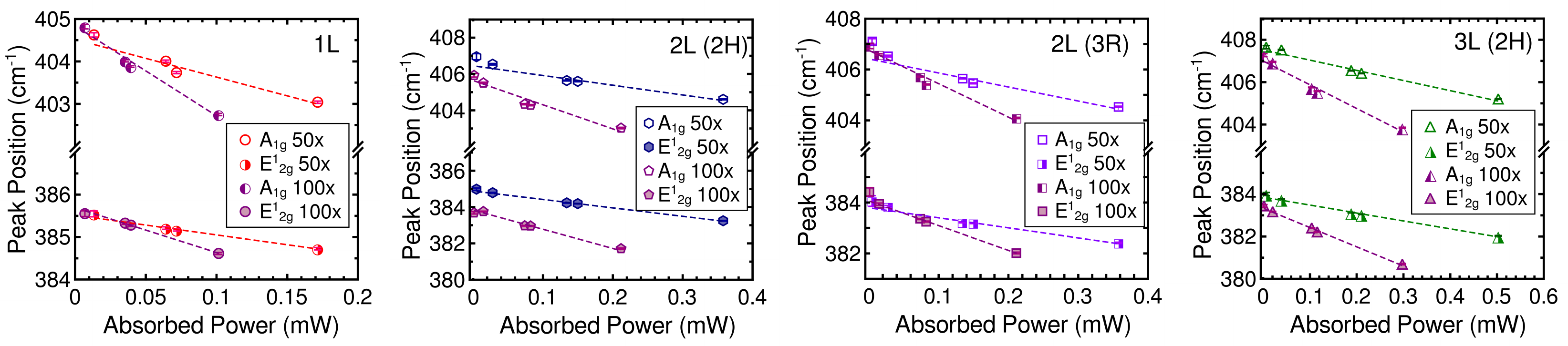}%
 \caption{\label{Figure4} (color online). The peak position of the A$_{1g}$ and  E$^{1}_{2g}$ phonon modes at different absorbed laser power for bi- and tri-layer MoS$_{2}$ crystals exfoliated on Si/SiO$_{2}$ substrate. (X and Y axis in linear scale). The dashed (-) lines represent linear fit to data points.}
 \end{figure*}
 
Since both the E$^1_{2g}$ and A$_{1g}$ modes have higher temperature coefficients, it allows the measurement of the local temperature rise of the supported MoS$_2$ crystals due to a change in incident laser power by employing changes in the phonon frequency. Raman measurements are performed with different laser power for two different spot-size objectives in ambient conditions. A 532 nm laser excitation is used with a step variable neutral density filter and 1800 lines/mm grating. The incident power is measured with a power meter, and the absorbed power is calculated by considering the layer number-dependent optical absorption for the 532 nm wavelength. The variation of the peak position of E$^1_{2g}$ and A$_{1g}$ phonon modes with absorbed laser power is displayed in Figure \ref{Figure4}. For all samples, the peak position redshifts with increased absorbed laser power. The fitting of data points with a linear function gives the power coefficient ($d\omega / dP=\chi_p$) of each Raman mode. The $\chi_p$ of the A$_{1g}$ mode appears more sensitive to variations in layer number than that of the E$^1_{2g}$ mode (Figure S3 \cite{SuppMat}). Since the A$_{1g}$ mode is insensitive to in-plane strain and its power coefficient is more sensitive to layer number, the temperature-induced softening of the A$_{1g}$ phonon mode is used to estimate local temperature rise and extract the thermal parameters. 

\begin{figure}[ht]
 \includegraphics[width=0.48\textwidth]{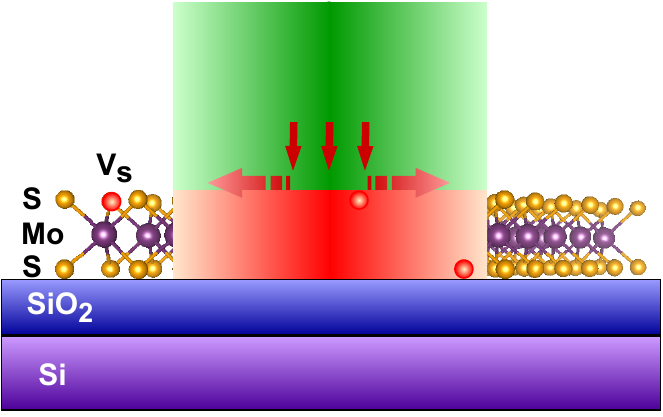}
  \caption{ Schematic of the laser heating and thermal transport in MoS$_2$ with sulfur vacancies exfoliated on Si/SiO$_2$ substrate.} \label{Figure5}
\end{figure}
 
Modeling of thermal transport in supported 2D materials subjected to laser heating has been reported previously.\cite{cai2010thermal,judek2015high,goushehgir2021simple} A laser beam with Gaussian intensity distribution is used for local heating of the sample, as shown in Figure 5. Under the steady-state condition, laser-induced heating is compensated by the heat flow towards the low-temperature heat sink. By assuming diffusive phonon transport, the heat conduction equation for a supported 2D material can be written as \cite{cai2010thermal}

\begin{equation}\label{Eq1}
\frac{1}{r}\frac{d}{dr}\biggl(r\frac{dT(r)}{dr}\biggr) - \frac{g}{\kappa t}(T(r)-T_{{a}}) + \frac{{q}_{0}^{\prime \prime}}{\kappa t }\exp\biggl(-\frac{r^{2}}{r_{0}^{ 2}} \biggr) = 0
\end{equation}

The temperature distribution $T(r)$ in the plane of 2D material has radial symmetry and Gaussian profile. Through out the thickness ($t$) of the 2D material same temperature profile is assumed. The layer thickness of mono-, bi- and tri-layer samples used for the calculations are 0.65 nm, 1.4 nm and 2.4 nm, respectively. The substrate is assumed to be at ambient temperature ($T_a)$ and acts as a heat sink.
Here, $ q_{0}^{\prime \prime} $ is the peak absorbed laser power per unit area at the center of the beam.
The absorbed laser power $Q={q}_{0}^{\prime \prime}\, \pi r_0^2 $ where $r_0 $ is the laser beam radius. Laser beam radius for 50x and 100x objectives is measured using Modified Knife edge method\cite{taube2015temperature} and the values are 0.495 $\mu$m and 0.297 $\mu$m, respectively. Details of laser beam size measurement is included in Section S6 of the supporting information.\cite{SuppMat} The optical absorption coefficients of mono-, bi- and tri-layer samples are taken as 5.8\%, 12.1\% and 17\%, respectively. The local temperature rise in the 2D crystal is defined as $  \Theta(r) = T(r) - T_{a} $, and the average temperature rise is given by 

\begin{equation}\label{Eq2}
\Theta_{{m}}(\kappa, g, r_0,Q) =  \frac{\int_{0}^{\infty}\Theta(r)\exp\biggl(-\frac{r^{2}}{r_{0}^{ 2}} \biggr)rdr }{\int_{0}^{\infty}\exp\biggl(-\frac{r^{2}}{r_{0}^{ 2}} \biggr)rdr }
\end{equation}

The thermal resistance which impedes the flow of heat in the system is obtained as $R_{m}=\Theta_{m}$/$Q$. The ratio of $R_{m}$ for two distinct laser beam sizes depends only on the material parameters $\kappa$ and $ g $. Hence on solving Equation (\ref{Eq1}) for two beam sizes, we are able to estimate  $\kappa$ and $ g $ uniquely. The estimated in-plane thermal conductivity and interface conductance values, along with the $R_{m}$ ratios, are presented in Table \ref{Table1}. Since the convection through air accounts for less than $0.15\%$ of the total heat conduction, its effect is ignored in the estimation of thermal parameters.\cite{zhang2015measurement} The in-plane thermal conductivity decreases as the layer number increases, whereas the total interface conductance per unit area across the 2D material-substrate interface increases. The reduction in in-plane thermal conductivity for increased layer number is primarily attributed to the intrinsic scattering mechanism of phonons.\cite{easy2021experimental} 

\begin{table}[ht]
	\caption{\label{Table1}{Thermal resistance ratio, in-plane thermal conductivity and interface thermal conductance (at room-temperature) of mono-, bi- and tri-layer MoS$_2$ crystals on $ \text{SiO}_2/ \text{Si}$ substrate}}
\begin{ruledtabular}
	\begin{tabular}{cccc}
	Sample &  $R_{m}$ ratio & $\kappa $ (W/mK) & $g $ (MW/m$^2$K)\\[0.2em]
	\hline
        1L &   2.282 $\pm$ 0.037 & 40 +8/-6 & 1.02 $\pm$ 0.03 \\
	2L (2H)  &   2.137 $\pm$ 0.084 & 33 +10/-7   & 1.09 +0.04/-0.05 \\
	2L (3R)  &  2.15 $\pm$ 0.13 & 32 +13/-10  & 1.07 +0.13/-0.15\\
	3L (2H) &  2.082 $\pm$ 0.068 & 28 $\pm$6 & 1.23 $\pm$0.05\\
    \end{tabular}
\end{ruledtabular}
\end{table}

In this work, the measured value of $\kappa $ of the monolayer MoS$_2$ sample is less than the previously reported value of 55 $\pm$ 20 W/mK in samples without any structural defects.\cite{zhang2015measurement} However, in the case of bilayer samples, our measured value of $\kappa $ is close to their reported value of 35 $\pm$ 7 W/mK.\cite{zhang2015measurement} In this context, it is essential to note that our MoS$_2$ samples contain sulfur vacancies which can alter their thermal transport properties. In our MoS$_2$ crystals, the additional surface vacancies created during annealing in an Ar/H$_2$ environment are mostly concentrated on exposed surfaces. Given the monolayer's high surface-to-volume ratio and larger exposed area, it contains the maximum defect density. Consequently, the highest reduction in thermal conductivity due to sulfur vacancies is observed for monolayer samples. Additionally, as the layer number increases, there might be a reduction in sulfur vacancy density in MoS$_2$ crystals. Hence, the observed lesser reduction in thermal conductivity of MoS$_2$ with sulfur vacancies in crystals with higher layer numbers could be due to their lesser defect density. Our observations align with recent theoretical predictions indicating that the presence of sulfur vacancies substantially reduces the thermal conductivity of monolayer MoS$_2$ primarily due to strong phonon localization and increased scattering by defects.\cite{polanco2020defect, wang2016remarkable} However, the impact of sulfur vacancies on the thermal conductivity of bulk MoS$_2$ is expected to be less significant due to the lower sulfur vacancy density in bulk crystals compared to monolayer crystals.\cite{polanco2020defect} Additionally, no significant variation is observed in the $\kappa $ and $g$ values of 2H and 3R stacked bilayer MoS$_2$ samples. It is essential to acknowledge that the specific impact of sulfur vacancies on thermal transport in MoS$_2$ can depend on additional factors such as vacancy type, concentration and distribution.\\

\section{Conclusions}

 In conclusion, we investigated the first-order temperature coefficients of interlayer and intralayer Raman modes of MoS$_2$ crystals with inherent sulfur vacancies exfoliated on Si/SiO$_2$ substrate. Later the $\kappa$ and $ g $ values of mono-, bi-, and tri-layer MoS$_2$ crystals are measured using Raman thermometry. The $\kappa$ value decreases, and the $ g $ value increases as the layer number increases. Importantly, we observed that the thermal conductivity is significantly suppressed by sulfur vacancies in monolayer MoS$_2$. However, this reduction in $\kappa$ becomes less evident as the layer number increases. Our results agrees with previous theoretical report that the suppression of thermal transport in MoS$_2$ by point defects is more prominent in monolayers and becomes negligible as layer number increases. This lesser reduction in thermal conductivity with increasing layer number could be attributed to the lower sulfur vacancy density at higher layer numbers. No significant variation is observed in the $\kappa$ and $ g $ values of 2H and 3R stacked bilayer MoS$_2$ samples.   

\begin{acknowledgments}
JB acknowledges the financial support from DST-SERB grant CRG/2020/000615. The authors thank the Central Instrumentation Facility (CIF) and Central Micro-Nano Fabrication Facility (CMFF), Indian Institute of Technology Palakkad and Central Instrumentation Facility (CIF), IISER Thiruvananthapuram for the experimental facilities. 
\end{acknowledgments}

%

\newpage
\section {Supplementary Information}
\title{Supporting Information \\~\\ Layer number and stacking order-dependent thermal transport in molybdenum disulfide with sulfur vacancies}
\maketitle

\section{S1. Layer number and stacking order}
Raman measurements are carried out in a HORIBA LabRAM HR Evolution Raman spectrophotometer setup using an Nd-YAG laser of wavelength 532 nm off-resonance excitation and grating of 1800 lines/mm. All the Raman modes are fitted with Lorentzian function to extract spectral parameters. The relative separation of E$^1_{2g}$ mode to A$_{1g}$ mode increases as the layer number increases for the first few-layers. 

\begin{figure}[ht]
    \renewcommand{\thefigure}{S1}
 \includegraphics[width=0.50\textwidth]{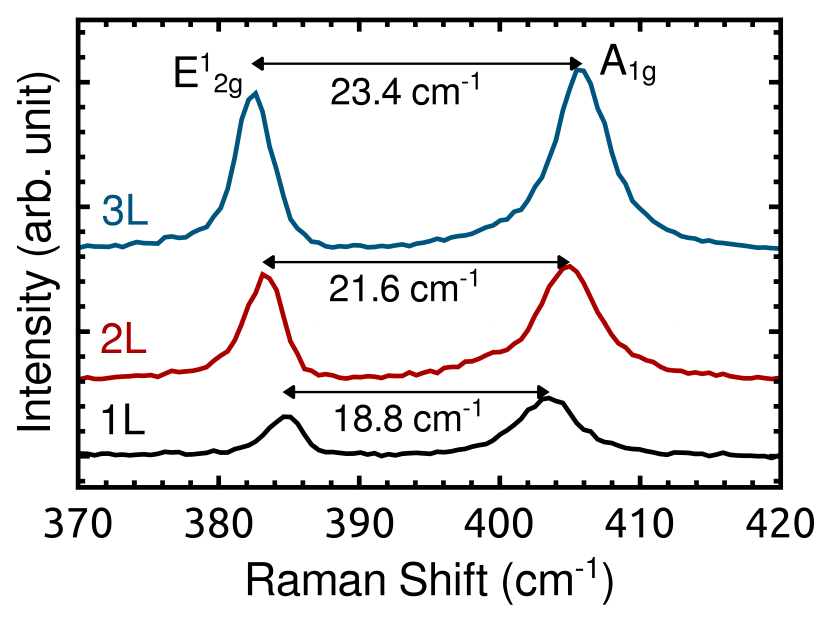}
  \caption{ High-frequency A$_{1g}$ and E$^1_{2g}$ Raman modes and their relative separation for mono-, bi- and tri-layer MoS$_2$\label{LayerNo}}
\end{figure}

\begin{figure}[ht]
     \renewcommand{\thefigure}{S2}
 \includegraphics[width=0.98\textwidth]{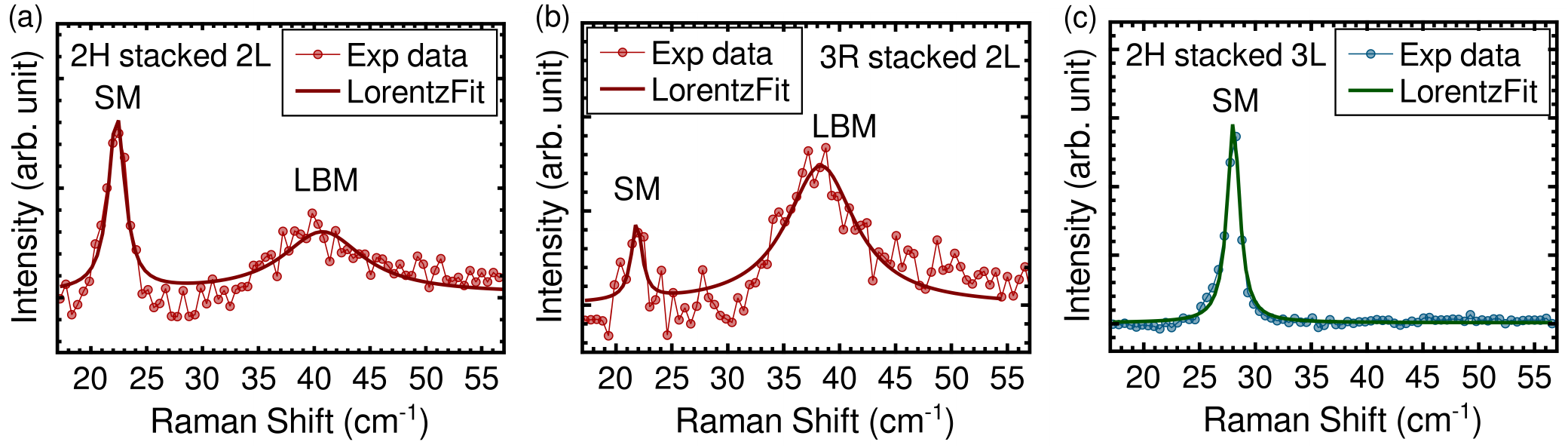}
  \caption{ Low-frequency shear mode and layer breathing mode of (a) 3R stacked bilayer (b) 2H stacked bilayer and (c) 2H stacked trilayer MoS$_2$\label{StackingOrder}}
\end{figure}

The position of shear mode for bilayer and trilayer MoS$_2$ is $\sim$22 cm$^{-1}$ and $\sim$28 cm$^{-1}$, respectively. The 3R polytype has lower frequency of the layer breathing mode compared to 2H polytype and is mainly due to the lower interlayer coupling strength in 3R stacking.\cite{van2019stacking} Additionally, the integrated intensity ratios of the layer breathing mode (LBM) to the shear mode (SM) strongly depends on the stacking order and interlayer interaction strength. I(LBM)/I(SM) of 3R polytype is nearly 5-6 times that of the 2H polytype. 


\section{S2. Raman measurements and thermometry}
 For all temperature-dependent measurements, excitation laser power is kept below 200 $\mu$W to avoid local heating. To perform measurements at higher temperatures a Linkam HFS600E-PB4 temperature-controller stage is used. Raman measurements are performed over the spectral range of 10-600 cm$^{-1}$. All laser power dependent measurements are carried out using a step variable neutral density filter at ambient conditions. 
\subsection{S2. 1. Temperature coefficient of low-frequency Raman modes}
The first-order temperature coefficients of low frequency Raman modes obtained for different samples are summarized in Table \ref{TempCoeffLFMs}. For trilayer samples both SM and LBM has the same phonon frequency. The temperature coefficient of layer breathing mode and higher orders of low-frequency modes of few-layer samples are not obtained due to less access to ultra-low frequencies.

\begin{table}[ht]
\renewcommand{\thetable}{SI}
	\caption{\label{TempCoeffLFMs}Temperature coefficients of Low-frequency interlayer modes of bi-, tri and few-layer MoS$_2$ crystals on $ \text{SiO}_2/ \text{Si}$ substrate}
\begin{ruledtabular}
	\begin{tabular}{ccc}
& \multicolumn{2}{c}{First-order temperature coefficient $\chi_T$, (cm$^{-1}$/K) } \\
	Sample &  Shear mode (SM) & Layer breathing mode (LBM)\\
	\hline
	2L (3R) MoS$_2$ & -0.0096 $\pm$ 0.0008 & -0.005 $\pm$ 0.002 \\
 	2L (2H) MoS$_2$ &  -0.0051 $\pm$ 0.0005 &  -0.005 $\pm$ 0.003 \\
	3L (2H) MoS$_2$ & -0.0041 $\pm$ 0.0006 &  - \\ 
 	FL MoS$_2$ & -0.0042 $\pm$ 0.0003 & - \\ 
    \end{tabular}
\end{ruledtabular}
\end{table}

\subsection{S2. 2. Laser power dependent Raman measurements}
\begin{figure}[ht]
    \renewcommand{\thefigure}{S3}
    \includegraphics[width=0.4\textwidth]{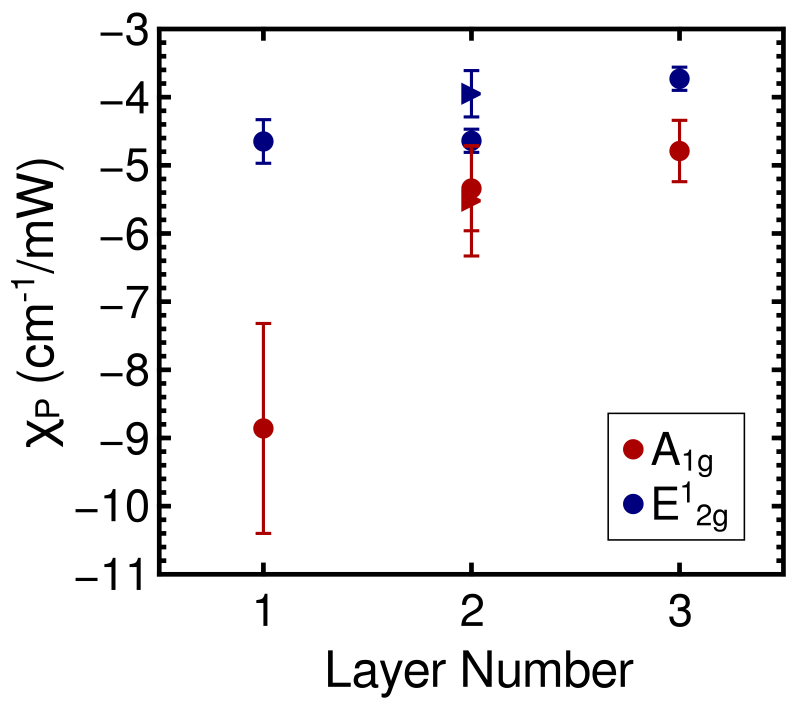}
  \caption{The value of power coefficient of E$_{2g}^1$ and A$_{1g}$ modes for mono-, bi- and tri-layer MoS$_2$ crystals on Si/SiO$_2$ substrate measured using 50x objective with 0.495 $\mu$m beam size. Triangles indicate 3R stacked bilayer. \label{PowerCoeffHFMs}}
\end{figure}

The power coefficients of E$^1_{2g}$ mode is less sensitive to variations in layer number. The magnitude of power coefficient of E$^1_{2g}$ and A$_{1g}$ Raman mode decreases with layer number as given in Figure \ref{PowerCoeffHFMs}. 

\section{S3. Estimation of $\kappa$ and $g$}

To determine the in-plane thermal conductivity ($\kappa$) and the interface conductance per unit area across the 2D material-substrate interface ($g$), we used the boundary conditions proposed by Cai et al.\cite{cai2010thermal} and solved Equation (2) in the manuscript using numerical integration; Estimated parameters are given in Table I in the manuscript.   

To confirm further thermal parameters are calculated by using the boundary conditions and analytical solution proposed by Goushehgir \cite{goushehgir2021simple}; Estimated parameters are given in Table SII. These values are in good agreement with the values obtained from numerical integration. When MoS$_2$ is supported on a substrate with a silicon dioxide (SiO$_2$) top layer, the interfacial thermal resistance at the MoS$_2$-substrate interface can become a significant factor in determining the overall thermal conductivity of the system. 

\begin{table}[ht]
\renewcommand{\thetable}{SII}
	\caption{\label{Table1}{In-plane thermal conductivity and interface thermal conductance (at room-temperature) of MoS$_2$ crystals}}
\begin{ruledtabular}
	\begin{tabular}{ccc}
	Sample &  $\kappa $ (W/mK) & $g $ (MW/m$^2$K)\\[0.2em]
	\hline
        1L &   40 +8/-6 & 1.02 $\pm$ 0.03 \\
	2L (2H)  & 33 +10/-7   & 1.09 +0.04/-0.05 \\
	2L (3R)  & 32 +13/-10  & 1.07 +0.13/-0.15\\
	3L (2H)& 28 $\pm$6 & 1.23 $\pm$0.05\\ 
    \end{tabular}
\end{ruledtabular}
\end{table}

\section{S4. Layer number and stacking order dependence of thermal properties}
\begin{figure}[ht]
     \renewcommand{\thefigure}{S4}
 \includegraphics[width=0.7\textwidth]{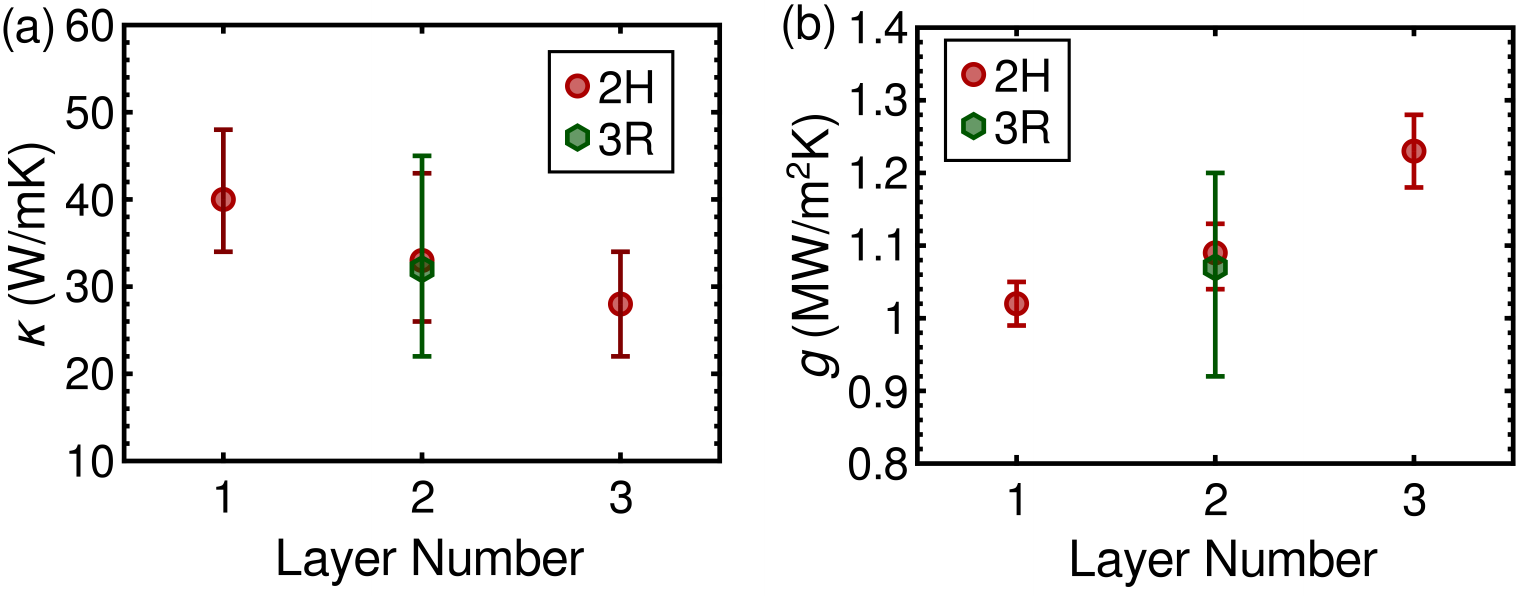}
  \caption{Variation of (a) in-plane thermal conductivity and (b) total interface thermal conductance of MoS$_2$ with layer number and stacking order \label{ThermalParameters}}
\end{figure}

Variation of thermal parameters with layer number and stacking order given in Figure \ref{ThermalParameters}.
\section{S5. Substrate heating}
\begin{figure}[ht]
   \renewcommand{\thefigure}{S5}
  \includegraphics[width=\textwidth]{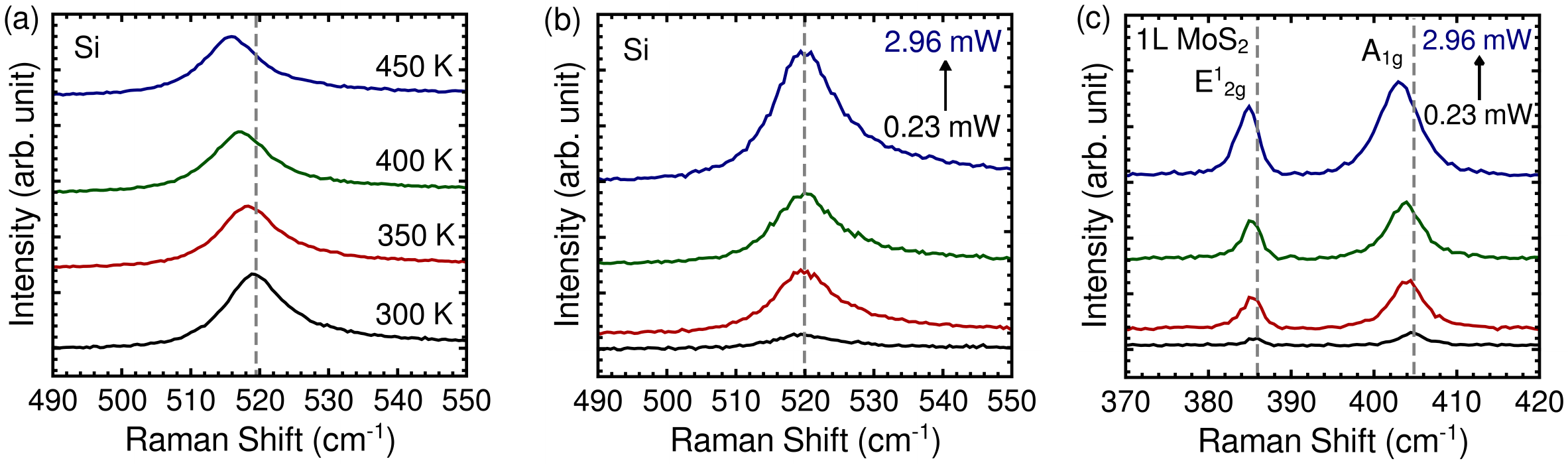}
  \caption{The variation of Silicon peak position with (a) temperature and (b) absorbed power for monolayer MoS$_2$ on Si/SiO$_2$ substrate. (c) The variation of high-frequency Raman modes of monolayer MoS$_2$ on Si/SiO$_2$ substrate. The power values given in the plots corresponds to the power absorbed by monolayer MoS$_2$.  \label{Si_Temp_power}}
\end{figure}

Variation of Si peak position with stage temperature and laser power is given in Figure \ref{Si_Temp_power} (a) and (b). No measurable shift in Si peak position is observed in the local laser heating. Due to high specific heat of SiO$_2$ layer it acts as a good heat sink and results in negligible heating of SiO$_2$ and underlying Si. 

\newpage
\section{S6. Laser beam size determination}
\begin{figure}[ht]
     \renewcommand{\thefigure}{S6}
 \includegraphics[width=0.7\textwidth]{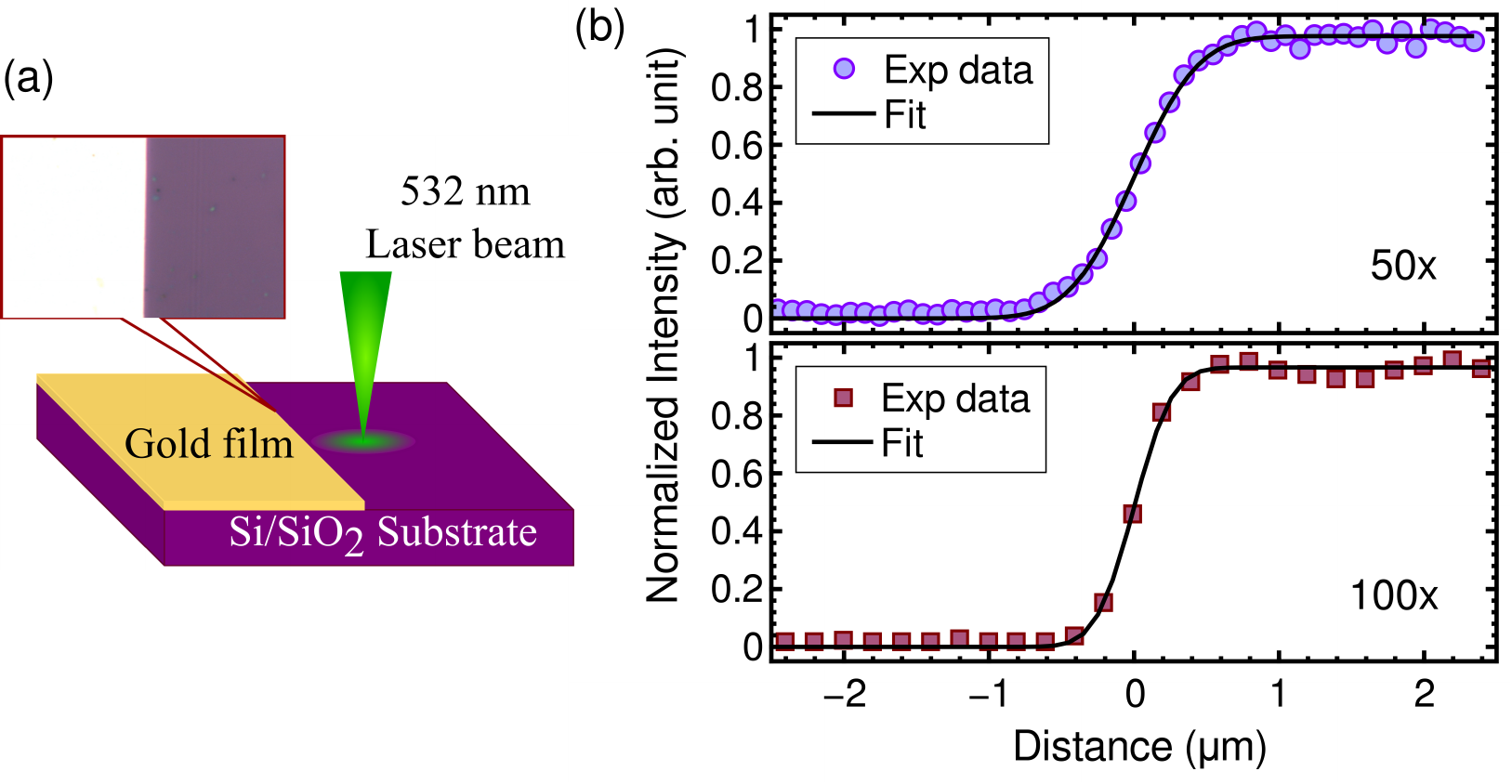}
  \caption{(a) Schematic of the laser beam size measurement set-up (optical image of the sharp edge given in inset). (b) The normalized intensity profile of the Si peak at different locations across the sharp edge measured using 50x and 100x objectives. The line represents the fit using Equation (\ref{eq_intensityfit}) \label{SpotSize}}
\end{figure}

The laser beam radius ($ r_0 $) is defined as the distance from the center of the beam at which intensity reduces to $1/e$ times its peak value. Laser beam size for both 50x and 100x objectives are measured by modified knife-edge method.\cite{taube2015temperature} A thin film of gold with sharp-edges is deposited on a Si/SiO$ _2 $ substrate. Linear Raman mapping is performed across the sharp edge of the gold film to detect the Raman signal of silicon. The acquisition time set as 2s, accumulations 2 and spectral range 420 cm$ ^{-1} $ to 600 cm$ ^{-1} $. Figure \ref{SpotSize} (a) shows the laser beam spot size measurement set-up and the optical image of the sharp edge. Figure \ref{SpotSize} (b) shows the normalized intensity profile of the Si peak at different locations across the sharp edge measured using 50x and 100x objectives. The normalized intensity profile is fitted with the function: 

 \begin{equation}\label{eq_intensityfit}
I(x) = \frac{I_0}{2}\biggl(1+erf\biggl(\frac{x-x_0}{w}\biggr)\biggr)
\end{equation}

The fitted normalized intensity profile is differentiated with respect to the distance x. The $ dI/dx  $ data has a Gaussian behavior with a functional form $ A . \exp{(-\frac{(x-x_0)^2}{w^2})} $. Since the source term in the heat conduction equation has the form $ \exp{(-\frac{r^2}{r_0^2})} $, the $w$ value can be identified as the laser beam radius, $ r_0 $. The measurements are repeated and the mean value for each objective is used for the estimation of thermal transport parameters.

\newpage

\section{S7. Transmission Electron Microscopy characterization}

\begin{figure}[ht]
    \renewcommand{\thefigure}{S7}
     \includegraphics[width=0.85\textwidth]{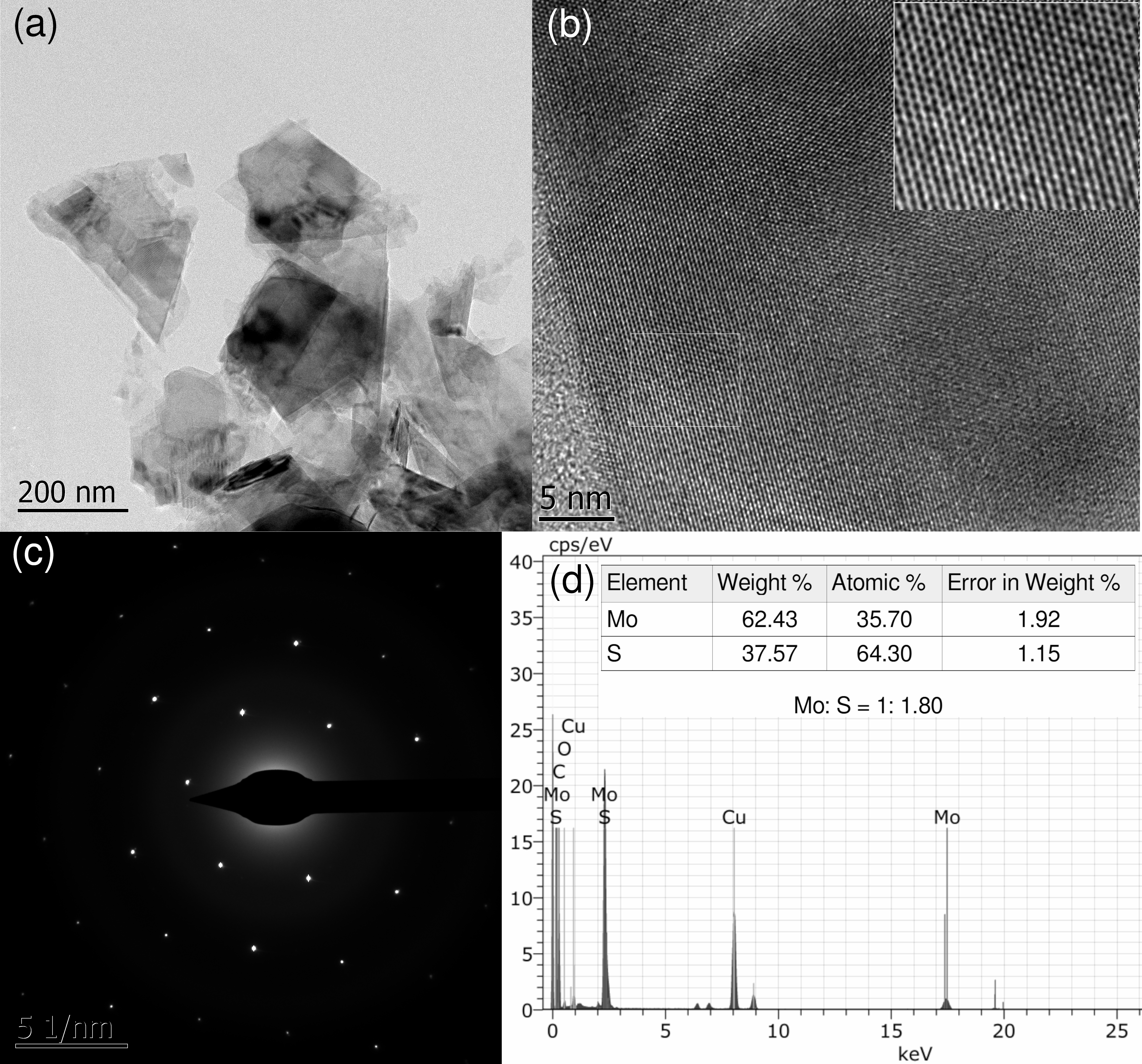}
     \caption{(a) The low-magnification image of the suspended MoS$_2$ flakes. (b) high resolution image: inset the zoomed image of the region in white rectangular box. (c) SAED pattern of the MoS$_2$ flake. The TEM-EDS spectrum and the atomic percentage of the Mo and S elements.}
     \label{fig:TEMresults}
 \end{figure}

Transmission Electron Microscopy (TEM) experiments were conducted on our MoS$_2$ samples before annealing to determine the inherent sulfur vacancy concentration. MoS$_2$ flakes were dispersed in ethanol and transferred to a carbon film supported copper grid. Figure \ref{fig:TEMresults}(a) shows a low-magnification image of the suspended flakes, while Figure \ref{fig:TEMresults}(b) presents a high-resolution image of a flake, with uniformly distributed bright spots corresponding to Mo atoms. The inset displays a zoomed image of the region in the white rectangular box. The Selected Area Electron Diffraction (SAED) pattern in Figure \ref{fig:TEMresults}(c) illustrates the high crystalline nature and trigonal prismatic coordination of the MoS$_2$ crystal. Additionally, the Transmission Electron Microscopy - Energy Dispersive Spectroscopy (TEM-EDS) spectrum in Figure \ref{fig:TEMresults}(d) and corresponding atomic percentages confirm the Mo:S ratio as 1:1.80, providing evidence for the presence of sulfur vacancies in the samples. The additional peaks for Cu, C, and O originate from the TEM grid.\\

\newpage

\end{document}